\documentclass[aps,twocolumn,amsmath,amssymb,floatfix,showpacs]{revtex4}

\usepackage{graphicx}% Include figure files
\usepackage{dcolumn} % Align table columns on decimal point
\usepackage{bm}      % Bold math
\usepackage{color}
\usepackage{ulem}

\renewcommand{\vec}{\mathbf}

\def\ADD#1{{\bf\textcolor{blue}{#1}}}         % addition
\def\ADsout#1{{\sout{\textcolor{blue}{#1}}}}       % change
\def\ADD#1{#1}         % addition
\def\ADsout#1{}       % change
\begin{document}

\title{High Reynolds number magnetohydrodynamic turbulence using a Lagrangian model}
\author{J. Pietarila Graham$^{1,2}$, P.D. Mininni$^{3,4}$, and A. Pouquet$^3$}
\affiliation{
$^1$Max-Planck-Institut f\"ur 
         Sonnensystemforschung, 37191 Katlenburg-Lindau, Germany \\
$^2$Department of Applied Mathematics \& Statistics, The Johns Hopkins University, 
         Baltimore, MD 21218, U.S.A. \\
$^3$Computational and Information Systems Laboratory, NCAR, 
         P.O. Box 3000, Boulder CO 80307-3000, U.S.A. \\
$^4$Departamento de F\'\i sica, Facultad de Ciencias Exactas y
         Naturales, Universidad de Buenos Aires and IFIBA, CONICET, 
         Ciudad Universitaria, 1428 Buenos Aires, Argentina.
             }
\date{\today}

\begin{abstract}
With the help of a model of magnetohydrodynamic (MHD) turbulence
tested previously, we explore high Reynolds number regimes up to
equivalent resolutions of $6000^3$ grid points in the absence of
forcing and with no imposed uniform magnetic field. For the given
initial condition chosen here, with equal kinetic and magnetic energy,
the flow ends up being dominated by the magnetic field, and the
dynamics leads to an isotropic Iroshnikov-Kraichnan energy
spectrum. However, the locally anisotropic magnetic field fluctuations
perpendicular to the local mean field follow a Kolmogorov law. We find
that the ratio of the eddy turnover time to the Alfv\'en time
increases with wavenumber, contrary to the so-called critical balance
hypothesis. Residual energy and helicity spectra are also considered;
the role played by the conservation of magnetic helicity is studied,
and scaling laws are found for the magnetic helicity and residual
helicity spectra. We put these results in the context of the dynamics
of a globally isotropic MHD flow which is locally anisotropic because
of the influence of the strong large-scale magnetic field, leading to
a partial equilibration between kinetic and magnetic modes for the
energy and the helicity.
\end{abstract}

\maketitle

\section{Introduction} \label{s:intro}

Magnetic fields are known to be dynamically important in a variety of
flows, for example \ADsout{in industrial installations for steel
manufacturing,} in the liquid core of the Earth, in stars and galaxies,
and they also play a role at the atomic level, e.g., in Bose-Einstein
condensates \cite{BEC_nature}. Their origin in astrophysics and
geophysics is still ill-understood, and their dynamics in the fully
nonlinear regime once the magnetic field has grown, possibly to
quasi-equipartition with the velocity, is ill-understood as
well. Numerous studies have shown, though, that magnetohydrodynamic
(MHD) turbulence differs in several ways from classical hydrodynamic
turbulence; for example, it is more intermittent than fluids
\cite{intermi1,intermi2,intermi-alpha}, more so in particular in two
space dimensions, and this intermittency may be either time-dependent
or dynamically constrained as observations of solar active regions
show \cite{abramenko}. Furthermore, numerical and theoretical
arguments indicate that MHD turbulence is less \ADD{spectrally} local than hydrodynamic
turbulence \cite{carati,alex} (see also
\cite{eyink,domaradzki,annrev}), and that MHD may lack universality in
its scaling properties \cite{dmitruk,muller,mason,ZZ}, including in
the absence of an imposed uniform magnetic field \cite{ed}. Finally,
the role of the correlation between the velocity and the magnetic
field has been known for a long time to be important, and this was
particularly emphasized when deriving an exact law involving linear
scaling with distance of third-order structure functions \cite{exact}
(see \cite{chandra} for the von K\`arm\`an relations for MHD in terms
of correlation functions).

Part of these differences may stem from the fact that, even in the incompressible MHD case, the system supports waves due to the effect of a large-scale magnetic field (hence the non-locality), whereas there are no such waves in incompressible hydrodynamics in the absence of external agents such as gravity or rotation. Of course, at small scales, plasma kinetic effects will come into play (e.g., in space plasmas), rendering the analysis more difficult. But even when one ignores such small-scale effects and concentrates on the dynamics of the large scales, one is still left with a variety of observations and of phenomenological approaches to the problem.

A wealth of new space data has come with the Cluster mission (which uses four identical spacecrafts to study the Earth's magnetosphere), and future projects such as the Magnetospheric Multiscale (MMS) mission will allow for more in-depth studies of MHD turbulence, albeit in a complex setting involving compressibility, boundary conditions, and geometrical effects as well as the aforementioned plasma phenomena. MHD turbulence can also be studied in the laboratory, although achieving high magnetic Reynolds numbers in liquid metals will remain a challenge for some time to come.

Of particular interest is how energy is distributed across scales in MHD turbulence. For the moment there is no clear-cut answer, with some indications that universality may be breaking \cite{ed}, but with the parameter(s) distinguishing between possible different classes of solutions not {well-}known. Direct numerical simulations (DNS) played an important role in these findings, as they allow for a rather controlled exploration of solutions, with well-defined parameters, but with the disadvantage that, in three space dimensions, the resolutions that are attainable with present-day computers are insufficient, by far, if one wants to model realistic geophysical and astrophysical flows with huge Reynolds numbers. Also, the computational cost of a detailed study of parameter space at sufficiently high resolution is currently out of reach.

One way out of this conundrum, in part, is to resort to modeling. There are numerous methods that have been devised over the years (see, e.g., recent reviews for fluids \cite{meneveau} and for MHD {\cite{jscicomp}).} Among  them, the Lagrangian averaged MHD model (LAMHD hereafter) developed in \cite{darryl} (see also {\cite{2d})} seems promising in that it allows to perform a quasi-DNS, in the sense that the Reynolds number is known and that the inertial range is extended, compared to a DNS performed on the same grid without the model, thanks to a different formulation of the equations that preserves the invariants, albeit in a different norm (see below).

In this context, we investigate properties of decaying MHD turbulence in the absence of an imposed uniform magnetic field using the LAMHD model. We thoroughly tested the LAMHD model previously, in two dimensions \cite{intermi-alpha, 2d, cancel} and in three dimensions \cite{3d}. Here we extend our computations to an equivalent grid resolution of the order of $6000^3$ points, with Taylor Reynolds number {at the maximum of dissipation of $\approx 2900$.}  The next section is devoted to definitions and the methodology, Sec.~\ref{s:simulation} gives details of the simulation, Sec.~\ref{s:global} discusses the global dynamics, and Sec.~\ref{s:alfv} discusses the role played by Alfv\'enic exchanges. Finally, Sec.~\ref{s:conclu} is our conclusion.

\section{Equations} \label{s:eq}
The MHD equations for an incompressible fluid with $\vec{v}$ and $\vec{b}$ respectively the velocity and magnetic fields in dimensionless Alfv\'enic units are:
\begin{eqnarray}
\partial_t\vec{v} + \boldsymbol{\omega} \times \vec{v} = \vec{j} \times \vec{b} - \boldsymbol{\nabla} p + \nu \nabla^2 \vec{v} \nonumber  \ , \\ 
\partial_t \vec{a} =  \vec{v} \times \vec{b} -\nabla \phi + \eta \nabla^2 \vec{a} \nonumber  \ , \\
\boldsymbol{\nabla} \cdot \vec{v} =  0 \ \ \ , \ \ \ \boldsymbol{\nabla} \cdot \vec{a} = 0,
\label{eq:mhd} \end{eqnarray}
with ${\bf b}=\nabla \times {\bf a}$, ${\bf a}$ being the magnetic vector potential in the Coulomb gauge. The potential $\phi$ and the pressure divided by the constant (unit) density $p$ are obtained self-consistently to ensure respectively the Coulomb gauge and the incompressibility of the velocity field. The kinematic viscosity is $\nu$ and the magnetic diffusivity $\eta$; $\boldsymbol{\omega} = \nabla \times \vec{v}$ is the vorticity, and ${\bf j}=\nabla \times {\bf b}$ is the current density. The magnetic Prandtl number $P_M=\nu/\eta$ is taken in what follows equal to unity.

In the absence of dissipation ($\nu=0\ , \eta=0$), the total energy
\begin{eqnarray}
E_T = E_v+E_b &=& \frac{1}{2} \left( ||{\bf v}||_2 + ||{\bf b}||_2 \right) = \nonumber \\
{} &\equiv & \frac{1}{2} \int \left(|{\bf v}|^2+ |{\bf b}|^2\right) d^3 x,
\end{eqnarray}
is conserved ($E_v$ and $E_b$ are respectively the kinetic and magnetic energy, and $||\,\cdot\,||_2$ denotes ${\cal L}_2$ norms). Two other quadratic quantities are preserved by the nonlinear terms in three dimensions (3D): the cross-correlation $H_C= \int {\bf v} \cdot {\bf b} \ d^3 x$, and the magnetic helicity $H_b= \int {\bf a} \cdot {\bf b} \ d^3 x$.

The Lagrangian averaged model introduces {\sl a priori} two filtering
lengths, $\alpha_{v,b}$, for the velocity and the magnetic fields in
such a way that the conservative structure of the equations is
preserved \cite{darryl}. Written in terms of the rough fields ${\bf
  v}$, ${\bf b}$, and ${\bf a}$, and of the filtered fields $\vec{u}$,
$\vec{{\cal B}}$ and $\vec{{\cal A}}$, the LAMHD equations read:
\begin{eqnarray} 
\partial_t\vec{v} + \boldsymbol{\omega} \times \vec{u} = \vec{j} \times \vec{{\cal B}} - \boldsymbol{\nabla} \Pi + \nu \nabla^2 \vec{v} \nonumber \\ 
\partial_t \vec{{\cal A}} =  \vec{u} \times \vec{{\cal B}} - \nabla \phi' + \eta \nabla^2 \vec{a} \nonumber \\
\boldsymbol{\nabla} \cdot \vec{v} =  \boldsymbol{\nabla} \cdot \vec{u} =  \boldsymbol{\nabla} \cdot \vec{a} =  \boldsymbol{\nabla} \cdot \vec{{\cal A}}  = 0,
\label{eq:lamhd} \end{eqnarray}
where $\Pi$ is a modified pressure, $\phi'$ a modified potential, and
with filtering being accomplished through normalized convolution
filters $L_{v,b}$ chosen to be the inverse of a Helmholtz operator,
namely $L_{v,b} = \mathcal H^{-1} = (1 -
\alpha_{v,b}^2\nabla^2)^{-1}$. The filtering lengths are related to
the ratio of the dissipation scales of the LAMHD flow and the modeled
MHD flow \cite{prandtl}; since we shall investigate flows with unit
magnetic Prandtl number, we take a common filter length for both the
velocity and magnetic field, $\alpha_{v,b}\equiv\alpha$. \ADD{That is,
  $\vec{u}=(1 -
  \alpha^2\nabla^2)^{-1}\vec{v}$ and $\vec{{\cal B}}=(1 -
    \alpha^2\nabla^2)^{-1}\vec{b}$.} The ideal quadratic invariants in
terms of ${\cal L}_2$ norms for MHD are transformed for LAMHD, being
expressed now in terms of $H_1^\alpha$ norms; this leads to
formulations that contain both the rough fields ${\bf v}$ and ${\bf
  b}$ and the filtered fields ${\bf u}$ and ${\bf {\cal B}}$, as can
be seen for example in the expression of the total energy:
\begin{eqnarray}
E_T^\alpha &=& \frac{1}{2} \left( ||u||^\alpha_1 + ||{\cal B}||^\alpha_1 \right)  \nonumber \\
{} &\equiv & \frac{1}{2} \int \left[ \left( \vec{u} - \alpha^2\nabla^2 \vec{u}\right)
      \cdot \vec{u} + \left(\vec{{\cal B}} - \alpha^2\nabla^2 \vec{{\cal B}}\right)
      \cdot\vec{{\cal B}}\right]~d^3x \nonumber \\
{} &=& \frac{1}{2} \int \left[\vec{v}\cdot\vec{u} + \vec{b}\cdot \vec{{\cal B}}\right]~d^3x.
\label{eq:alpha_inv}
\end{eqnarray}
Similarly, the magnetic helicity $H_b^{\alpha}=\int {\bf a} \cdot {\bf{\cal B}} \ d^3 x$ in its $H^\alpha_1$ norm is preserved by the LAMHD equations. A form of cross-correlation is also preserved, and the LAMHD equations also have the equivalent of an Alfv\'en flux-conservation theorem (based on these equivalences, and for the sake of clarity, the superscript $\alpha$ will be suppressed in the sections presenting numerical results unless strictly required).

The Reynolds number in the following is defined as
\begin{equation}
Re=U_{rms}L_0/\nu \ ,
\end{equation}
with $U_{rms}$ and $L_0$ the r.m.s.~velocity and the integral scale respectively. The Taylor Reynolds number $R_{\lambda}=U_{rms}\lambda/\nu$ is defined using either the kinetic or magnetic Taylor scale:
\begin{equation}
\lambda_{v,b} = 2\pi \sqrt{E_{v,b}^{\alpha}/\Omega_{v,b}^\alpha} \ ,
\end{equation}
where $\Omega_{v}^\alpha=||\boldsymbol{\omega}||_1^\alpha$ and $\Omega_b^\alpha=\Omega_b=||{\bf j}||_2$ are respectively the enstrophy and square current (or magnetic enstrophy) in LAMHD, proportional respectively to the kinetic and magnetic dissipation in the model. Taylor wavenumbers are defined as $k_{{\lambda}_{v,b}}=2\pi/\lambda_{v,b}$.

The numerical implementation of the LAMHD equations is performed using a pseudo-spectral code for which inversion of the Helmholtz operator is straightforward. The code is parallelized using MPI (Message Passing Interface) and has been tested in a variety of conditions (see, e.g., \cite{gomez}); a hybrid version of the code, using OpenMP as well, allows efficient parallelization for higher resolutions using a larger number of processors \cite{hybrid}. The computational box has a size of $(2\pi)^3$, and wavenumbers vary from $k_{min}=1$ to $k_{max}=N/3$ using a standard 2/3 de-aliasing rule, $N$ being the number of grid points per direction.

\section{Simulation set-up} \label{s:simulation}
The dynamics of MHD turbulence is quite complex and there may be different regimes, as characterized for example by their energy spectra, either isotropic or anisotropic depending on the presence of an imposed uniform magnetic field ${\bf b}_0$. With ${\bf b}_0\equiv 0$ and no forcing, a high resolution DNS using $1536^3$ grid points \cite{pablo_1536} showed evidence of isotropic Iroshnikov-Kraichnan (hereafter, IK) energy scaling \cite{IK}. More recently, it was shown \cite{ed} that energy spectra may differ measurably even when global quantities ($E_T$, $E_b/E_v$, $H_b$, and $H_C$) are the same at the initial time: IK-like, weak turbulence like (hereafter, WT) \cite{galtier}, or Kolmogorov-like spectra (hereafter, K41) \cite{K41} were obtained, the discriminating factor being the ratio of Alfv\'enic time $\tau_A$ to the eddy turn-over time $\tau_{NL}$. These times are defined respectively as:
$$ \tau_A =\ell/b_0 \hskip0.2truein ,   \hskip0.2truein  \tau_{NL} = \ell/v_{\ell} \ , $$
$v_{\ell}$ being the characteristic velocity at scale $\ell$; in the absence of an imposed mean field, $b_0$ in the above expression is taken as the amplitude of the magnetic field at the gravest mode.

In this paper, we examine one of these cases further by performing computations at higher Reynolds number using the LAMHD model, resulting in an equivalent grid of roughly $6000^3$ points and a Taylor Reynolds number of $\approx 2900$. The case of interest is the one that gives IK-like energy scaling, and we therefore consider only one initial condition, similar to the one studied in Ref.~\cite{pablo_1536} with $1536^3$ spatial resolution. By ``equivalent grid,'' we mean the following: a comparison with the $1536^3$ DNS with $\eta=\nu=2\times10^{-4}$ was carried solving the LAMHD equations on grids of $256^3$, $384^3$, and $512^3$ points to validate the results of the model \cite{jscicomp,3d}. The runs presented here, with {$1024^3$} grid points and viscosity and magnetic diffusivity $\eta=\nu=5\times10^{-5}$, would require a $\approx 6000^3$ resolution in a DNS.

The filter length $\alpha$ is chosen to be $4\pi/k_{max}$, i.e., the filter wavenumber is $k_{max}/2$, and the grid resolution is $\Delta x=2\pi k_{max}^{-1}/3$ (see, e.g., \cite{3d} for details on how the filter length is chosen for a given resolution). The initial conditions are close but not identical to those used in \cite{pablo_1536}; they consist of a superposition of Arn'old-Beltrami-Childress (ABC) flows distributed in the first four shells ($k\in [1,4]$), with super-imposed noise with a wide spectrum as in \cite{pablo_1536}. However, the noise has a steeper energy distribution, $\propto k^{-6}$, instead of $\propto k^{-3}$.  The choice of a shallower distribution of random noise in \cite{pablo_1536} was guided by the desire to reach the maximum of dissipation in a short computational time; the computational constraint not being so strong in the LAMHD model, we resort to a steeper initial random distribution so as to let turbulent spectra evolve in a less constrained manner, allowing us to examine in this paper as well whether qualitative differences in behavior take place or not under such a change.

The steeper distribution of noise here also results in slightly different values for the relative initial helicities (when compared with the run in \cite{pablo_1536}), defined respectively as
$$\rho_v=H_v/\sqrt{E_v \Omega_v} \hskip0.2truein , \hskip0.2truein \rho_b =H_b/\sqrt{E_a E_b} \ .$$
Here, $E_a$ is the ${\cal L}_2$ norm of the vector potential, and $H_v=\int {\bf u} \cdot {\bf \omega} \, d^3x$ is the kinetic helicity, which measures deviations from mirror symmetry in the hydrodynamic flow. Because of Schwarz inequality, we have $|\rho_{v,b}| \le 1$.

The resulting flow at $t=0$ has $E_v=E_b=1$, $\rho_v=0.78$, and $\rho\ADD{_b}=0.94$, compared to $E_v=E_b=1$, $\rho_v=0.49$, and $\rho\ADD{_b}=0.54$ for the DNS run in \cite{pablo_1536}, leading to stronger nonlinearities and faster initial evolution in the previous work. Note that at $t=0$ we can use either ${\cal L}_2$ or ${\cal H}_1^\alpha$ norms for the LAMHD energy and other global quantities, since they differ only after the 4th digit.

\section{Global dynamics} \label{s:global}
\subsection{Time evolution}

\begin{figure}
\hskip-0.2truein
\includegraphics[width=8.6cm]{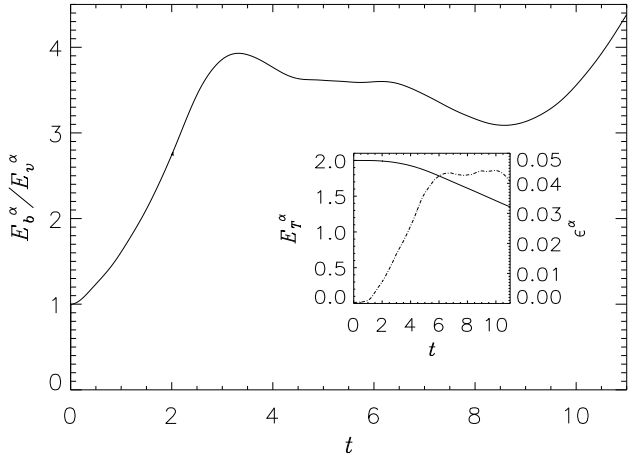}
\caption{Ratio of total magnetic energy $E_b^\alpha$ to kinetic energy $E_v^\alpha$ as a function of time, and (inset) total energy, $E_T^\alpha=E_v^\alpha+E_b^\alpha$ (solid line), and total dissipation, $\epsilon^\alpha = \nu\Omega_v^\alpha+\eta\Omega_b$ (dash-dotted line) versus time. Note that the superscript $\alpha$ is suppressed in the labels, and will be suppressed in later figures for clarity.}
\label{fig:emevst} \end{figure}
\begin{figure}
\includegraphics[width=8.6cm]{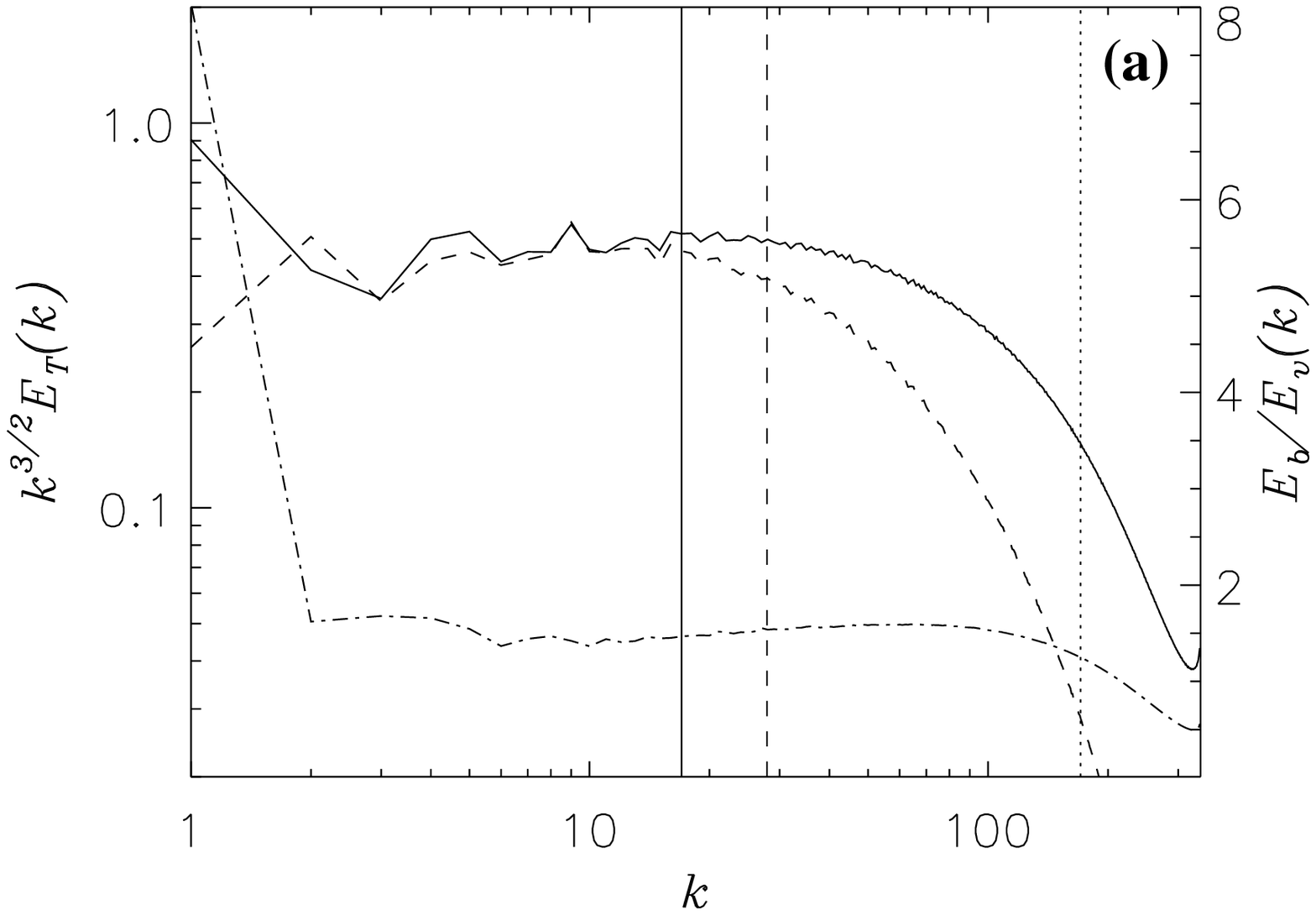}
\includegraphics[width=8.6cm]{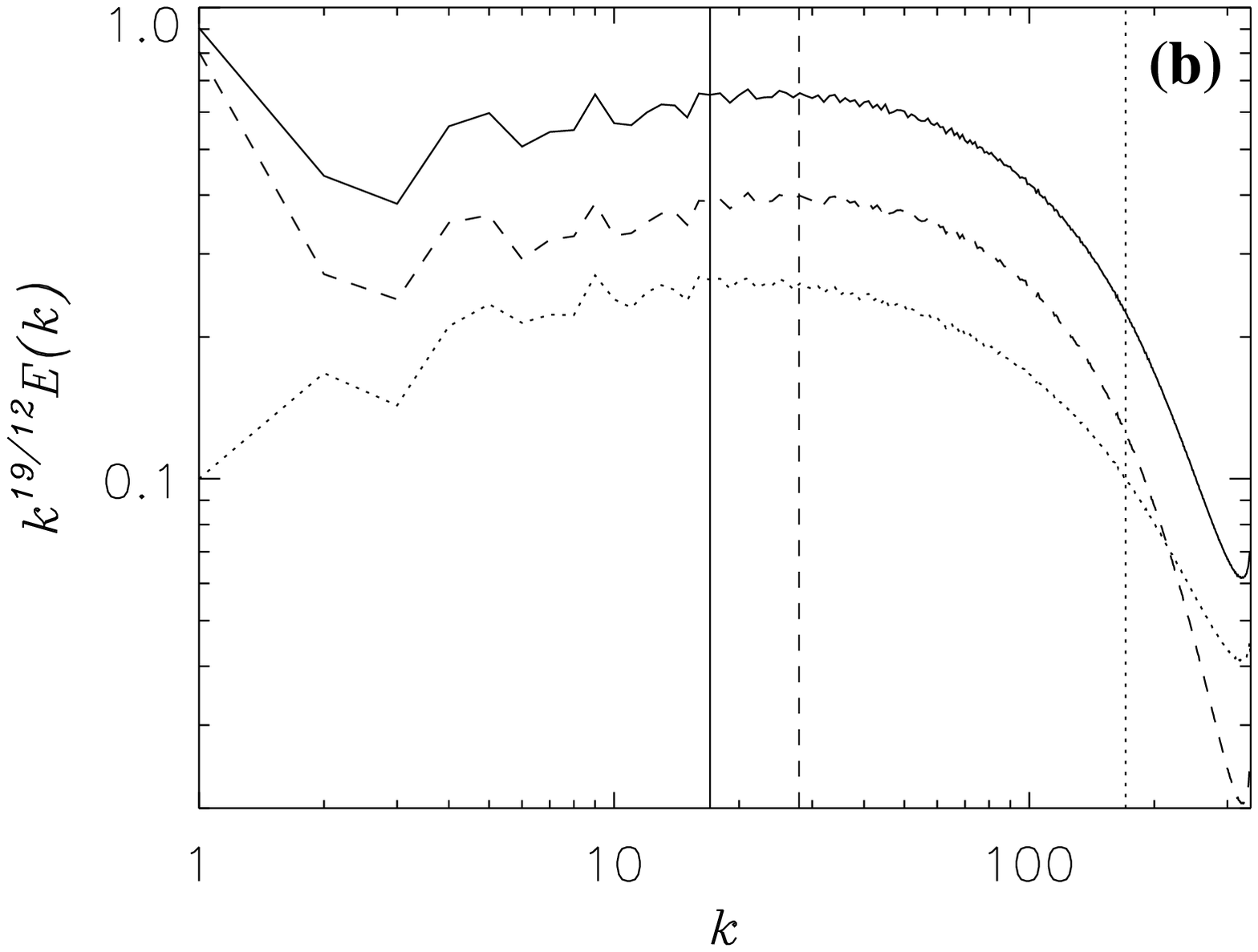}
\caption{\ADD{\it (a):} $E_T(k)$ averaged over time with $t\in[6.5,11]$, compensated by $k^{3/2}$ (solid line, left scale) and ratio of magnetic to kinetic energy as a function of $k$ (dash-dotted line, right scale).
\ADD{Also shown is $E_T(k)$ from a lower $Re$ $1536^3$ DNS \cite{pablo_1536}
(dashed line) demonstrating the efficacy of the model.}
Here and in all subsequent figures, the vertical solid, dashed, and dotted
lines denote respectively the wavenumbers corresponding to the kinetic
Taylor scale, the magnetic Taylor scale, and the filtering scale $\alpha$
(beyond which the model is expected to differ from MHD \ADD{with $\eta=\nu=5\times10^{-5}$}).
\ADD{\it (b):} Total (solid line), magnetic (dashed line) and kinetic (dotted line) energy spectra compensated by $k^{19/12}=k^{(3/2+5/3)/2}$ (see text for details). The IK slope ($-3/2$) is favored against the K41 ($-5/3$) slope until the magnetic Taylor scale.}
\label{fig:spectra} \end{figure}

As expected from the larger values of relative helicities, development of a fully developed turbulent regime takes a longer time in the present run compared to the DNS run on a grid of $1536^3$ points. Indeed, the maximum of total dissipation for the LAMHD flow is reached for $t_{\ast}\approx 6.5$ (as opposed to $\approx 3.8$ for the $1536^3$ DNS run) as can be seen in Fig.~\ref{fig:emevst} (inset). The total dissipation remains approximately constant between the maximum and the final time of the computation, $t\approx 11$, providing us with a quasi-steady state at small scales for several turn-over times. Except for this delay in the onset of the formation of small scale gradients, both flows evolve in similar manners. Figure 1 also shows the temporal evolution of the total energy, and of the ratio of magnetic to kinetic energy. The magnetic energy grows rapidly at the expense of its kinetic counterpart, and the ratio between these two quantities then fluctuates around a value close to $\approx 3.5$ for times between $\approx 3$ and $\approx 10$, with a tendency for further increase toward the end of the run.

\subsection{Energy scaling}
During the time of approximately constant dissipation rate, the energy spectrum presents a clear IK law, as can be seen in Fig.~\ref{fig:spectra}\ADD{(a)}, which displays the total energy spectrum compensated by $k^{3/2}$. In this figure, and in the subsequent ones, the vertical solid line represents the classical Taylor scale based on the velocity field, the vertical dash line the magnetic Taylor scale, whereas the vertical dotted line indicates the filter wavenumber $2\pi/\alpha$ (beyond which the model is expected to differ from MHD). Note that no bottleneck is observed, i.e., there is no enhancement of the spectrum at the onset of the dissipative range, as already observed in several MHD simulations. The IK law seems a good approximation to the dynamics in the range of wavenumbers from $k\approx 4$ to $k\approx 2\pi/\lambda\ADD{_b}\approx 28$, i.e., all the way to the magnetic Taylor scale. 

To verify this, Fig.~\ref{fig:spectra}\ADD{(b)} also shows the total, kinetic, and magnetic energy spectra compensated by $k^{19/12}$. The spectral index $19/12$ corresponds to the mean value between the IK and K41 spectral indices, i.e., $(3/2 + 5/3)/2$. As a result, a positive slope in the compensated plot indicates the spectrum is closer to IK, while a negative slope indicates the spectrum is closer to K41 scaling. The total and magnetic energy spectra monotonously increase up to the magnetic Taylor wavenumber, and only the kinetic energy spectrum flattens and has a small negative slope between the kinetic and magnetic Taylor wavenumbers.

Figure \ref{fig:spectra}\ADD{(a)} also gives (with scale on the right) the ratio $E_b^\alpha(k)/E_v^\alpha(k)$; it appears remarkably flat starting at the beginning of the IK inertial range, and it also displays a slight enhancement peaking around $k=80$ (associated with the faster decay of the kinetic energy spectrum after the kinetic Taylor scale). The approximate constancy, with $E_b(k)\approx 1.4\ E_v(k)$, denotes a partial Alfv\'enization of the flow in most of the inertial range, but with here at $k = k_{min}$ a magnetic field which can be evaluated as being roughly 3 times larger than the large scale velocity field, and at least 10 times larger than the r.m.s.~value of the turbulent fluctuations.

Note that, at scales smaller than the filter $\alpha$, the balance between the velocity and the magnetic field for the model is inverted, with a slight excess of kinetic energy (e.g., note $E_b^{\alpha}(k)/E_v^{\alpha}(k)<1$ in this range). This may be attributed to an effective hyper-resistivity in LAMHD for the magnetic field in the sub-filter scales; at these scales, the model differs from MHD. Indeed, in the DNS reported in \cite{ed}, this decrease of $E_b/E_v$ was only seen close to the dissipation frequency and the magnetic to kinetic ratio remained above unity at all scales.

\begin{figure}
\includegraphics[width=8.6cm]{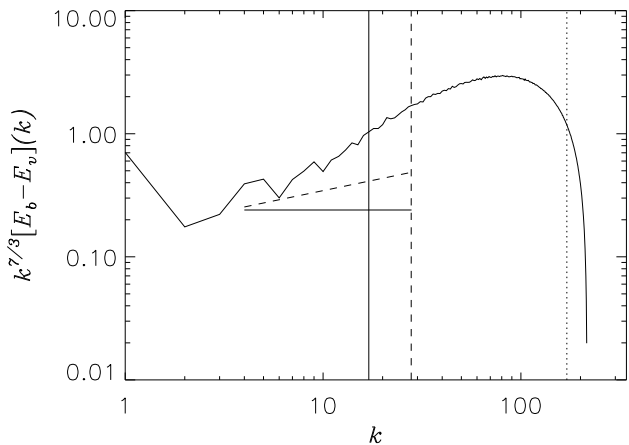}
\caption{Residual energy spectrum {$E_R(k)=E_b(k)-E_v(k)$} averaged in
  the same time interval as in Fig.~\ref{fig:spectra} \ADD{and compensated
  by $k^{7/3}$.} The dashed line
  indicates a slope of $-2$ up to the Taylor magnetic scale, while the
  solid \ADD{horizontal} line indicates a slope of $-7/3$; the best fit from $4\le k\le
  28$ gives $-1.4\pm0.1$.}
\label{fig:residual} \end{figure}

The residual energy spectrum
\begin{equation}
E_R(k)=E_b^\alpha(k)-E_v^\alpha(k) ,
\end{equation}
is shown in Fig. ~\ref{fig:residual}. 
{A heuristic argument indicates that}
 for K41 scaling, the residual energy is expected to vary as $k^{-7/3}$,
 while for IK scaling the residual energy should go as $k^{-2}$
 \cite{strong}. In the inertial range of our LAMHD run, the slope is closer
 to (although shallower than) $-2$ \ADD{(a result also seen in Ref. \cite{HB06}.) T}his point is further discussed in Sec.~V.

\begin{figure}
\includegraphics[width=8.6cm]{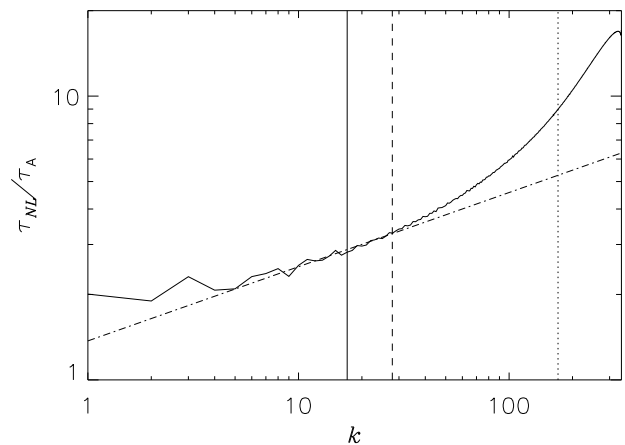}
\caption{Ratio of the eddy turn-over time $\tau_{NL}$ to the Alfv\'en time $\tau_A$ as a function of wavenumber averaged in the same interval of time as in Fig. \ref{fig:spectra} (solid line); the power law fit (dot-dash line) has an exponent of $0.3$ with an error of $\pm 0.1$.}
\label{fig:ratio} \end{figure}

We show in Fig.~\ref{fig:ratio} the ratio $\tau_{NL}(k)/\tau_A(k)$ as a function of wavenumber, where the eddy turnover time and the Alfv\'en time are defined as
\begin{equation}
\tau_{NL}(k) = \frac{1}{k\sqrt{kE_v^\alpha(k)}} ,
\label{eq:tnl}
\end{equation}
\begin{equation}
\tau_A(k) = \sqrt{\frac{2}{k^2 E_b^\alpha(k=1)}},
\end{equation}
respectively.

Since $E_b^{\alpha}(k)/E_v^{\alpha}(k)\sim$ constant (see Fig.~\ref{fig:spectra}\ADD{(a)}), this implies that the associated characteristic time ratio $\tau_A/\tau_{NL}$ must be a function of wavenumber. Indeed, the ratio is clearly not constant as the wavenumber increases, and seems to follow in the inertial range a $k^{0.3\pm 0.1}$ power law (Fig.~\ref{fig:ratio}). If we assume $E_v(k)\sim E_T(k)\sim k^{-3/2}$, this leads to $\tau_{NL}\sim k^{-3/4}$ and to the ratio shown in Fig.~\ref{fig:ratio} varying as $k^{1/4}$. When we assume K41 scaling with $E_v(k)\sim k^{-5/3}$, the ratio of times varies as $k^{1/3}$. Finally, for WT we would have $k^{1/2}$. The first two solutions are compatible with the data, in particular considering we are not taking into account  that (i) intermittency corrections to the spectral indices can alter these indices, and (ii) the kinetic and total energy spectra may not follow the same power law. Indeed, one should remember that there is no constraint on the behavior of $E_{v,b}(k)$ separately, since the invariant that cascades directly is the total energy. In fact, the stronger constraint on the spectral behavior of MHD fields comes from the exact law in terms of the energy and energy fluxes of Els\"asser variables, or in terms of the total energy and cross helicity and their fluxes (as well as for magnetic helicity). These laws imply complex correlations between the velocity and the magnetic field \cite{exact}, but do not imply that power laws must be followed by each individual field. Finally, note that this result is not in agreement with the critical balance hypothesis \cite{GS} which postulates $\tau_{NL}(k)=\tau_A(k)\sim k^{-1}$; in fact, solving this relationship for the kinetic energy spectrum $E_v(k)$ this hypothesis would lead to $E_v(k)\sim k^{-1}$ in the isotropic case, implying a logarithmic divergence which is unlikely (the case with an external uniform magnetic field may be different).

\begin{figure}
\includegraphics[width=8.6cm]{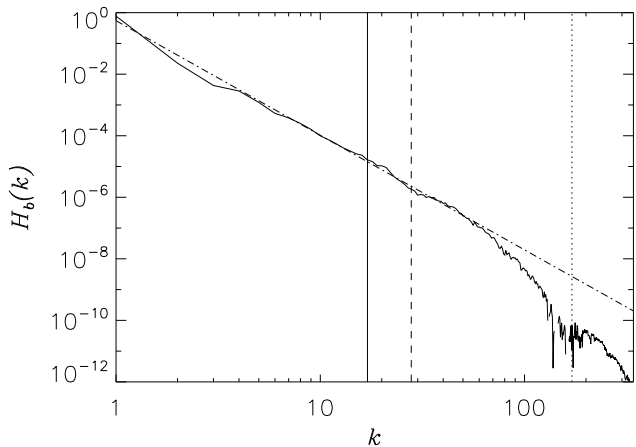}
\caption{Spectrum of magnetic helicity averaged over the same interval of time as in Fig. \ref{fig:spectra} (solid line); the dot-dashed line indicates a power-law fit with exponent $-3.7$ and an error of $\pm0.1$.
% Note that $H_b>0$ always,
%It is negative only for $k=131,141,142,144,145, 160-162,164,339-341$. 
%(whereas $H_v$, not shown,  is negative for $k<38$ and $k>323$).
} \label{fig:mhel} \end{figure}

\subsection{Helicity scaling}

Examining now the magnetic helicity spectrum, we observe that it shows
a single scaling in the energy containing range, in the inertial
range, and beyond the Taylor scales (in contrast to the spectra
discussed before). It follows rather closely a $k^{-3.7\pm0.1}$ law,
as can be seen in Fig.~\ref{fig:mhel}. Such a steep spectrum for $H_b$
has been observed before \cite{intermi2,malapaka}. The slope does not
correspond to the phenomenological analysis done in \cite{strong}
under the hypothesis of an inverse cascade of magnetic
helicity. Indeed, in this simulation all the helicity is concentrated
initially in $k \in [1,4]$, and therefore there is no separation
between the helicity containing scale and the size of the box for an
inverse cascade of magnetic helicity to develop. The power law in
$H_b(k)$ seems therefore to be associated with a transfer of the
helicity towards smaller scales by the direct energy cascade. As will
be discussed in the next section, this transfer is required to satisfy
a balance imposed by \ADD{Alfv}\'enization and magnetic field
induction.  \ADD{Note that in \cite{BrSu2005}, such a steep spectrum for
  $H_b(k)$ is interpreted as a Kolmogorov spectrum for both the
  current helicity and the kinetic helicity, through sweeping by the
  large scales.  }

\begin{figure}
\includegraphics[width=8.6cm]{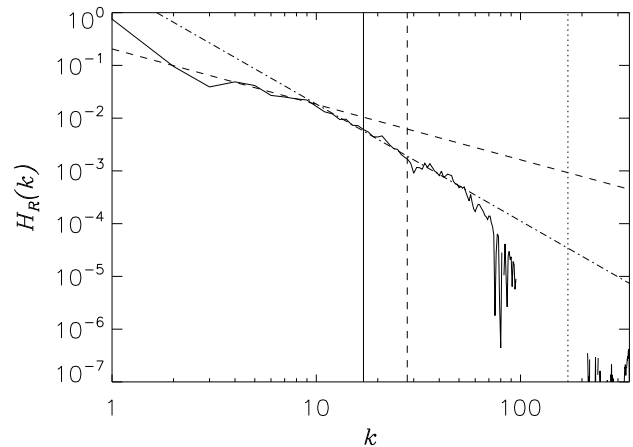}
\caption{Residual helicity $H_R(k)=k^2H_b(k)-H_v(k)$. The dashed line indicates a power law $\sim k^{-1}$, while the dash-dotted line indicates $\sim k^{-2.2}$.}
\label{fig:reshel} \end{figure}

Similarly to the case of energy, a residual helicity spectrum can be defined as
\begin{equation}
H_R(k)=k^2H_b(k)-H_v(k) .
\end{equation}
The factor $k^2$ multiplying $H_b(k)$ gives the current helicity spectrum, which is dimensionally equivalent to $H_v(k)$. This functional is known to play an important role in the nonlinear stage of the dynamo \cite{strong}. Fig.~\ref{fig:reshel} shows $H_R(k)$ in the simulation. Two ranges become apparent, with a knee in the spectrum around $k\approx9$. For larger wavenumbers, the residual helicity seems to follow a $\sim k^{-2.2}$ law. We observe that $H_b(k)>0\ \forall k$, and $k^2H_b(k)-H_v(k)$ is positive until $k \approx 100$.

\subsection{Local anisotropy}

\begin{figure}
\includegraphics[width=8.6cm]{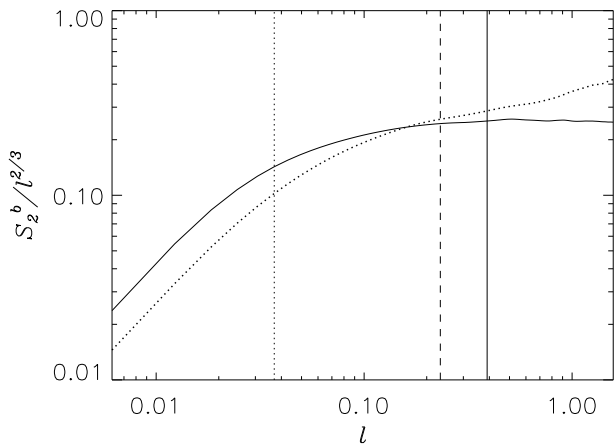}
\caption{Second-order longitudinal structure function (for $t=7$) of the magnetic field, compensated by a Kolmogorov law $l^{2/3}$, for perpendicular (solid line) and parallel (dotted line) increments with respect to the local mean magnetic field. The structure function for parallel increments (dotted line) displays, at scales larger than the Taylor scales, a scaling $\sim l_\parallel$.}
\label{fig:scaling} \end{figure}

We further examine in Fig.~\ref{fig:scaling} the local scaling of the magnetic field when expressed in terms of its second-order longitudinal structure function, separated into perpendicular and parallel spatial increments,
\begin{equation}
S_2^b(l_{\perp,\parallel}) = \left< \left\{ \left[ {\bf b}({\bf x} + {\bf l}_{\perp,\parallel}) - {\bf b}({\bf x}) 
    \right] \cdot \frac{{\bf l}_{\perp,\parallel}}{|{\bf l}_{\perp,\parallel}|} \right\}^2 \right> .
\end{equation}
Here, $\perp$ and $\parallel$ denote space-varying directions with respect to the direction of a local mean magnetic field ${\bf b_{0,loc}}$, defined as an average of ${\bf b}$ in a box of edge length $\pi$ (the flow integral scale is $L_0\approx 2.9$) around each point ${\bf x}$ for which the data is sought. Therefore, these structure functions do not measure global anisotropy (the flow is globally isotropic) or global scaling, but rather local anisotropy and scaling with respect to a local mean magnetic field.

The points at which these correlation functions are computed are chosen at random, and the figure is the result of an analysis performed for  {$2.5 \times 10^7$} points. Noteworthy is the variation with scale of the ratio of the perpendicular to the parallel component, with a cross-over near the magnetic Taylor scale. At smaller scales, $S_2^b(l_\perp)> S_2^b(l_\parallel)$, and the ratio increases as the scale decreases until reaching the filter scale. This is in sharp contrast with hydrodynamical turbulence, including in the presence of an imposed rotation: isotropy is recovered at small scale in such flows because the ratio of the inertial wave time to the eddy turn-over time gets larger as the wavenumber grows, or in other words the influence of anisotropy is felt at large scale but becomes secondary at small scales. It is also noteworthy that the local perpendicular magnetic field seems to display a scaling closer to K41 turbulence, while the parallel increments, having $S_2^b(l_\parallel) \sim l_\parallel$, display a scaling close to $\sim k_\parallel^{-2}$.

\section{Alfv\'enic exchanges}\label{s:alfv}

Is there a dynamical role for Alfv\'enic exchanges beyond a tendency toward equipartition? Even if the spectral behavior of the total energy can be obtained from first principles and phenomenological considerations as to what is the relevant timescale for energy transfer (the eddy turn-over time, the Alfv\'en time, or their combination in a transfer time $\tau_{tr}=\tau_{NL}^2/\tau_A$ that embodies the slowing down of nonlinear coupling in the presence of waves \cite{IK}), the behavior of individual spectra is not so straightforwardly determined. For the kinetic and magnetic energy spectra, there are indications of differing spectral behavior both in the solar wind \cite{podesta} and in numerical simulations \cite{pablo_1536b}. For the magnetic helicity, flows forced at intermediate scales have been observed to undergo an inverse cascade to large scales including in the supersonic case (see \cite{balsara} and references therein), but in the decaying case and with initial conditions at large scale, the behavior of $H_b(k)$ does not appear to be universal  \cite{malapaka}. In this section, we discuss the scaling laws observed in our simulations in terms of Alfv\'enic exchanges and magnetic induction.

We start with the residual energy $E_R(k)$, Fig. \ref{fig:residual}. The way to derive $E_R(k)\sim k^{-2}$ is to suppose that \cite{strong}: (i) the residual energy is small compared to the total energy, in a ratio $(\tau_A/\tau_{NL})^2$, i.e.,
\begin{equation}
E_R(k)\sim (\tau_A/\tau_{NL})^2 E_T(k) \ ;
\label{ER}
\end{equation}
and (ii) that the kinetic energy spectrum appearing in $\tau_{NL}$ (see Eq.~\ref{eq:tnl}) behaves in the inertial range as the total energy. Assuming further that $E_T$ follows an IK $k^{-3/2}$ law leads to $E_R(k)\sim k^{-2}$, but the second hypothesis may not be fulfilled in all cases \cite{podesta,pablo_1536b} and indeed we find a shallower residual spectrum, $E_R\sim k^{-3/2}$.

The understanding of the origin of the discrepancies in the scaling of magnetic helicity, Fig. \ref{fig:mhel}, is partial at best. We focus here on the spectral scaling in the direct cascade energy range presented in the previous section, as opposed to the inverse cascade of magnetic helicity which is not accessible in our simulations. It is advocated in \cite{malapaka} that the scaling can be attributed to a combination of factors involving the competing nonlinearities in MHD turbulence, modeled using the two-point closure developed in \cite{strong}. A phenomenological way to recover the balance advocated in \cite{malapaka} is to assume partial Alfv\'enization of the flow, leading to a quasi-equipartition of the velocity and magnetic field, both for the symmetric part (energies) and the anti-symmetric part (helicities) of their respective correlation tensors. In other words, one could expect to have, in the order of an Alfv\'en time (based on the large-scale magnetic field), the following balance
\begin{equation}
E_v(k)\sim E_b(k) \ \ , \ \ H_v(k) \sim k^2 H_b(k) \ .
\label{eq:equipar1}\end{equation}
Note however that equipartition is not achieved exactly, as observed in the solar wind. One can then postulate that this incomplete equipartition be equally occurring for the energy and for the helicity. That is, that the ratio
\begin{equation}
\frac{E_v(k)/E_b(k)}{H_v(k) /\left[k^2 H_b(k)\right]}
\label{eq:equipar2}\end{equation}
be close to a constant in the inertial range. This condition is approximately fulfilled in our simulation as well as in the DNS in \cite{pablo_1536,pablo_1536b}.

The relationship in Eq.~(\ref{eq:equipar2}) can be derived in a different manner, that takes into account the dynamics of the magnetic field and further clarifies how the system can evolve towards such balance. To do this, we take a mean-field approach to induction processes in the equation for the magnetic field, associated in the mechanically forced case with the growth of a seed magnetic field by dynamo action. By splitting the magnetic field into ${\bf b} = {\bf b_0} + {\bf b}'$, where ${\bf b_0}$ is a large-scale magnetic field that grows under the influence of the small-scale kinetic helicity, and ${\bf b}'$ are the turbulent magnetic fluctuations, one obtains under the mean-field approximation:
\begin{equation}
\frac{\partial {\bf b_0}}{\partial t} = \alpha^{\prime} \nabla \times {\bf b_0} .
\label{eq:meanfield}
\end{equation}
Here $\alpha^{\prime}\propto - \tau_{corr} \left<H_v\right>_{ss}$, where $\left<.\right>_{ss}$ indicates averaging over the small-scale turbulent helical field, and $\tau_{corr}$ is a velocity field correlation time \cite{SKR} (see \cite{brandenburg} for a review). Taking the dot product of the above equation with the large-scale magnetic potential ${\bf a}_0$ and integrating over space, one gets the following evolution equation for the large-scale magnetic helicity:
\begin{equation}
\frac{d}{dt} \int{ {\bf a}_0 \cdot {\bf b}_0 d^3 x} = 2 \alpha^{\prime} \int{ |{\bf b}_0|^2 d^3 x} .
\label{eq:meanfield2}
\end{equation}
Note that dissipation has been neglected in these equations, and that the large-scale magnetic energy ($\int |{\bf b}_0|^2 d^3 x$) appears in the equation upon integration under suitable boundary conditions.

As a result of Eq.~(\ref{eq:meanfield2}), the helical velocity field injects helicity into the large-scale magnetic field. Since total magnetic helicity is conserved in the ideal case, the opposite amount of magnetic helicity must be created at small scales, i.e., the following dynamical balance must hold independently of the scale (with a change in sign for large and small scales):
\begin{equation}
\frac{d H_b}{dt} \sim \pm\tau_{corr} H_v E_b  \ .
\label{eq:equipar3}\end{equation}
The dynamical balance can be understood using the conceptual ``stretch, twist, and fold'' (STF) dynamo \cite{Vainshtein72}. Each time a closed magnetic flux tube is twisted by the helical velocity field, smaller-scale magnetic field lines are twisted in the opposite direction, thus pushing towards small scales some magnetic helicity. This process removes magnetic helicity from the large scales and allows the magnetic field to ``disentangle'' through reconnection events, destroying in that way magnetic helicity \citep{Alexakis06,annrev}.

The correlation time $\tau_{corr}$ can be taken to be the eddy
turn-over time, $\tau_{corr} \sim \tau_{NL}$,
since the induction effect given by Eq.~(\ref{eq:equipar3}) is
associated with deformation of magnetic field lines by turbulent
eddies.  From
Eq.~(\ref{eq:equipar3}), assuming again that the temporal evolution
follows the eddy turn-over time $\tau_{NL}$, i.e., by approximating
$d/dt \sim 1/\tau_{NL}$, we find
\begin{equation}
H_b\propto H_v E_b/ (k^2 E_v)
\label{eq:HM} \end{equation}
from which Eq.~(\ref{eq:equipar2}) follows.
These phenomenological considerations show that the regeneration of the magnetic field by dynamo action is in fact compatible with an Alfv\'enic balance, and that these two processes may be occurring on similar time scales in MHD turbulence. In a decaying flow such as here, there may not be sufficient time available to see such Alfv\'enic and dynamo exchanges take place repeatedly, but it would be of interest to examine the helicity exchanges in a forced flow; this may provide a geometrical origin of \ADD{Alfv}\'enization, through the dynamic alignment of relevant vectors, as is known to occur on a nonlinear time scale \cite{matthaeus}.

The constraint given by Eq.~(\ref{eq:HM}) is compatible with the spectrum of magnetic helicity observed in Fig.~\ref{fig:mhel}. In hydrodynamic turbulence, $k^{-1}H_v/E_v \sim 1/k$ \cite{Chen}, and under this assumption $H_b \sim k^{-2} E_b$ (in this simulation, the ratio $k^{-1}H_v/E_v$ decays slightly faster with $k$). For $E_b \sim E_T \sim k^{-3/2}$ or steeper, then $H_b \sim k^{-7/2}$ or steeper. The $k^{-3.7\pm0.1}$ law in Fig.~\ref{fig:mhel} seems therefore consistent with a transfer of the helicity towards smaller scales by the direct energy cascade, to satisfy the balance imposed by \ADD{Alfv}\'enization or induction.

We finally consider the spectrum of residual helicity from the point of view of \ADD{Alfv}\'enization. To derive its scaling, we assume that, similarly to the residual energy case, we have:
\begin{equation}
H_R(k)\sim (\tau_A/\tau_{NL})^2 H_C(k) \ .
\label{HR}
\end{equation}
The cross-correlation appears here for at least two reasons: On the one hand, helicity is a pseudo-scalar and thus the r.h.s.~of Eq.~(\ref{HR}) must involve a pseudo-scalar as well. On the other hand, it is known that total energy and cross-correlation play inter-linked roles in the dynamics of MHD flows. This is particularly striking when writing the exact laws in MHD turbulence that stem from conservation of $E_T$ and $H_C$ \cite{exact}, as mentioned before. As a result, the phenomenological formulation in Eq.~(\ref{HR}) seems the most plausible.

If we now assume that $H_C(k)$ scales as the total energy, i.e. $H_C(k)\sim E_T(k) \sim k^{-3/2}$,  then Eq.~(\ref{HR}) yields $H_R(k) \sim k^{-2}$. However, it should be noted that the spectrum of cross-helicity in MHD turbulence may be steeper than IK, as, e.g., $H_C(k)\sim k^{-2}$, a spectral law which was obtained in two-point closures \cite{grappin}. In this case the spectrum leads to $H_R(k) \sim k^{-5/2}$. For $9\le k\le50$, we find $H_R(k)\sim k^{-2.2}$ which does not allow us to decide between the two solutions. Moreover, note that the correlation coefficient is very low in this computation, of the order of $\rho_C = H_C/E_T \approx 1.5 \times 10^{-3}$, with a spectrum strongly fluctuating with wavenumber.

These simple arguments indicate that there may be in fact dynamical constraints on the behavior of the flow that link the velocity and magnetic field statistics, both for the energy and the helicity, and that go beyond a strict derivation of scaling laws in turbulence based on the conservation of a sole invariant, e.g., the magnetic helicity or the total energy.

\section{Discussion and conclusion} \label{s:conclu}

In this paper we presented results from numerical simulations of MHD turbulence in the absence of external forcing, using a Lagrangian averaged model to attain the largest possible Reynolds number at a given spatial resolution. At very high Reynolds number, the separation of scales between the energy-containing range and the dissipation range is huge and, with a power-law decrease of the distribution of energy among scales, for all practical purposes the large-scale magnetic field acts as a locally uniform field for the small scales; with one important difference though: that there is no imposed direction to the flow and that it is in fact isotropic at large scales (no $k=0$ component). In other words, we can say that we have global isotropy and homogeneity. This is the case in the Solar Wind at very large scale: the locally uniform field in one of the spiral arms, changes direction with the spiral arm, and globally can be viewed as isotropic. However, at small scale, the effect of the large-scale field is such that the dynamics is anisotropic, or more precisely, locally anisotropic.

As a result of the global isotropy in our run, the total energy spectrum develops an inertial range compatible with Iroshnikov-Kraichnan phenomenology, and other spectral quantities are consistent with this scaling. Moreover, the small scales are seen to go through an Alfvenization process that can be attributed to a local mean magnetic field.

However, magnetic helicity presents a wide inertial range with a spectrum steeper than what usual phenomenological arguments predict. What then is governing the behavior of magnetic helicity? In the inverse cascade range, a dimensional Kolmogorov-like analysis leads to $H_b(k)\sim k^{-2}$ \cite{strong}, and numerical data using the integro-differential MHD equations in the framework of the Eddy-Damped Quasi-Normal Markovian (EDQNM) closure does corroborate the existence of such a regime. One should note that the argument in \cite{strong} pre-supposes the independence of the cascades that may take place in MHD turbulence, that of total energy to small scales, that of cross-helicity likely to small scales as well, and that of magnetic helicity to large scales. However, the equipartition relationships in Eqs.~(\ref{eq:equipar1}), (\ref{eq:equipar2}), and (\ref{eq:HM}), link the different correlators that one can build on the physical variables in a much more intricate way. We do know already that there are other such complex relationships involving coupling of these quantities, namely the exact laws that stem from the conservation of $E_T$, $H_C$ and $H_b$ \cite{exact,exact_HM}.

 It is advocated in \cite{malapaka} that, in the decay and the forced case, different dynamics may take place. It is true that for a steady-state inverse cascade to develop, separation of scales and forcing are in principle required, and the present simulation has initial conditions with most of the magnetic helicity concentrated at the largest scale in the box. At this point, it is therefore important to make a distinction between the behavior of magnetic helicity in its inverse cascade range, and its behavior in the direct cascade range of energy as observed in our simulation. In the direct cascade range, helicity seems to be transported to smaller scales by the energy cascade, and its spectrum can be explained from phenomenological arguments that take into account Alfv\'enization of the flow, or equivalently, the dynamics of magnetic field induction leading to Alfv\'enization.

These arguments are however dependent on the spectrum of the kinetic and magnetic energy separately, on the spectrum of kinetic helicity $H_v$, and on the spectrum of $H_C$. Indeed, the exact relationship for magnetic helicity derived in \cite{exact_HM} is more complex in its structure than the laws for the other invariants; it involves the fields themselves, not structure functions (based on field differences). Moreover, the electromotive force ${\bf v}\times {\bf b}$ (associated with the magnetic induction) plays a central role (the law in \cite{exact_HM} derives only from the induction equation). Furthermore, the third-order correlators of third-order tensors involving pseudo-vectors have a much more complex structure than for the full symmetric (non-helical) case, since it requires {\it a priori} four scaling functions to define the properties of the dynamics. Other examples where the presence of helicity leads to new scaling laws are known \cite{mininni_helrot}. Thus, we can expect, in a similar fashion, that there could be different forms of the magnetic helicity spectrum in MHD depending on, e.g., the kinetic helicity or the intensity of the cross-helicity in the flow.

\ADD{These helicity effects open} the door to a lack of universality for spectra of invariants (see also \cite{yousef,ed,malapaka}), since different regimes may occur in different MHD flows. Note that these arguments differ from the critical balance regime for MHD turbulence advocated in \cite{GS} and which does not seem to be observed in high Reynolds number MHD, at least in the globally isotropic and decaying case (see \cite{ed}, and Fig. \ref{fig:ratio}). In spite of the different possible scaling laws arising for magnetic helicity, it would be interesting to know if all satisfy the condition given by Eq. (\ref{eq:HM}). We know of at least a few more examples satisfying this condition, e.g., the flow in Ref. \cite{pablo_1536b}, and the flows studied in the pioneering work of M\"uller and Malapaka \cite{malapaka}.

Finally, and as mentioned in the introduction, one other peculiarity of MHD turbulence is the fact that interactions are less \ADD{spectrally} local than in hydrodynamic turbulence. This has been known for a long time in its simplest version, that of Alfv\'en waves. The large-scale magnetic field is responsible for the propagation of waves that put in equipartition the small-scale velocity and magnetic fields; this is the basis for the modification to a Kolmogorov energy spectrum as proposed in \cite{IK}. In fact, non-local effects have been quantified in MHD turbulence (see \cite{alex,annrev} and references therein). Nonlocality may also break down the underlying hypothesis leading to $H_b(k)\sim k^{-2}$ even for the inverse cascade range, and may intermingle small-scale velocity and magnetic excitations with the large-scale magnetic field. As stressed in \cite{yousef}, nonlocality breaks the simple self-similarity known to occur in fluids at least at second order; it is also shown in \cite{yousef} that the magnetic field at small scale may have a folded structure that is consistent with the exact law derived in \cite{exact}. Thus, the exact MHD laws, and their associated balances presented here, which are derived independently of locality assumptions, may therefore prove useful to reach a better understanding of MHD turbulence.
\vskip0.5truein

\begin{acknowledgments}
We acknowledge useful discussions with W.C. M\"uller and S.K. Malapaka. Computer time was provided by NCAR and by the Rechenzentrum Garching (RZG) of the Max Planck Society and the Institut f\"ur Plasmaphysik. NCAR is sponsored by the National Science Foundation. PDM acknowledges support from UBACYT grant No. 20020090200692, PICT grant No. 2007-02211, and PIP grant No. 11220090100825. 
\end{acknowledgments}

%\bibliography{phd}

\end{document}